\documentclass{ecai} 

\usepackage{latexsym}
\usepackage{amssymb}
\usepackage{amsfonts}
\usepackage{amsmath}
\usepackage{amsthm}
\usepackage{booktabs}
\usepackage{enumitem}
\usepackage{graphicx}
\usepackage{color}
\usepackage{graphicx} 
\usepackage{algorithm}
\usepackage{algorithmic}

\usepackage[utf8]{inputenc} 
\usepackage[T1]{fontenc}    
\usepackage{url}            
\usepackage{booktabs}       
\usepackage{amsfonts}       
\usepackage{nicefrac}       
\usepackage{microtype}      
\usepackage{xcolor}         
\usepackage{multirow}%
\usepackage{tabularray} 
\usepackage{amsfonts}
\usepackage{amsmath}
\usepackage{fancyhdr}
\usepackage{graphicx}
\usepackage{caption}
\usepackage{subcaption}
\usepackage{caption}
\usepackage{listings}
\usepackage{tabularx}

\usepackage{newfloat}
\usepackage{listings}

\newcommand{\BibTeX}{B\kern-.05em{\sc i\kern-.025em b}\kern-.08em\TeX}
\newcommand{\overbar}[1]{\mkern 1.5mu\overline{\mkern-1.5mu#1\mkern-1.5mu}\mkern 1.5mu}

\begin{document}


\begin{frontmatter}



\title{TRADES: Generating Realistic Market Simulations 
\\
with Diffusion Models}


\author[A]{\fnms{Leonardo}~\snm{Berti}\thanks{Corresponding Author. Email: berti.1883894@studenti.uniroma1.it. Work done during M.Sc. at Sapienza University of Rome.}}
\author[B]{\fnms{Bardh}~\snm{Prenkaj}}
\author[A]{\fnms{Paola}~\snm{Velardi}} 

\address[A]{Sapienza University of Rome}
\address[B]{Technical University of Munich}

\paperid{5552}

\begin{abstract}
Financial markets are complex systems characterized by high statistical noise, nonlinearity, volatility, and constant evolution. Thus, modeling them is extremely hard. Here, we address the task of generating realistic and responsive Limit Order Book (LOB) market simulations, which are fundamental for calibrating and testing trading strategies, performing market impact experiments, and generating synthetic market data. We propose a novel \textbf{TRA}nsformer-based Denoising \textbf{D}iffusion Probabilistic \textbf{E}ngine for LOB \textbf{S}imulations (\textbf{TRADES}). TRADES generates realistic order flows as time series conditioned on the state of the market, leveraging a transformer-based architecture that captures the temporal and spatial characteristics of high-frequency market data. There is a notable absence of quantitative metrics for evaluating generative market simulation models in the literature. To tackle this problem, we adapt the predictive score, a metric measured as an MAE, to market data by training a stock price predictive model on synthetic data and testing it on real data. We compare TRADES with previous works on two stocks, reporting a $\times 3.27$ and $\times 3.48$  improvement over SoTA according to the predictive score, demonstrating that we generate useful synthetic market data for financial downstream tasks. Furthermore, we assess TRADES's market simulation realism and responsiveness, showing that it effectively learns the conditional data distribution and successfully reacts to an experimental agent, giving sprout to possible calibrations and evaluations of trading strategies and market impact experiments. To perform the experiments, we developed DeepMarket, the first open-source Python framework for LOB market simulation with deep learning. In our repository, we include a synthetic LOB dataset composed of TRADES’s generated simulations. 
\end{abstract}

\end{frontmatter}


\section{Introduction}
A realistic and responsive market simulation\footnote{To abbreviate, throughout this paper, by ``market simulation'', we refer to a Limit Order Book (LOB) market simulation for a single stock.} has always been a dream in the finance world \cite{raberto2005modeling,levy2000microscopic,mizuta2016brief,jacobs2004financial}.  Recent years have witnessed a surge in interest towards deep learning-based market simulations \cite{li2020generating, nagy2023generative, cont2023limit, coletta2021towards}.  An ideal market simulation should fulfill four key objectives: {(1)} enable the calibration and evaluation of algorithmic trading strategies including reinforcement learning models; {(2)} facilitate counterfactual experiments to analyze {(2.1)} the impact of orders \cite{webster2023handbook}, and {(2.2)} the consequences of changing financial regulations and trading rules, such as price variation limits;  {(3)} analyze market statistical behavior and stylized facts in a controlled environment; {(4)} generate useful granular market data to facilitate research on finance and foster collaboration.

For these objectives, two key elements are paramount: the {\textit{realism}} and the {\textit{responsiveness}} of the simulation. Realism refers to the similarity between the generated probability distribution and the actual data distribution. Responsiveness captures how the simulated market reacts to an experimental agent's actions.  Furthermore, the generated data \textit{usefulness} is crucial for achieving the last objective (4). Usefulness refers to the degree to which the generated market data can contribute to other related financial tasks, such as predicting the trends of stock prices \cite{prata2023lobbased, sirignano2019deep}. 

Backtesting and Interactive Agent-Based Simulations (IABS) \cite{byrd2020abides} are two of the most used traditional market simulation methods. Backtesting assesses the effectiveness of trading strategies on historical market data. It is inherently non-responsive since there is no way to measure the market impact of the considered trading strategies, making the analysis partial. 
IABS, on the other hand, enables the creation of heterogeneous pools of simplified traders with different strategies, aiming to approximate the diversity of the real market. However,  obtaining realistic multi-agent simulations is challenging, as individual-level historical data of market agents is private, which impedes the calibration of the trader agents, resulting in an oversimplification of real market dynamics.

Recent advances in generative models, particularly Wasserstein GANs \cite{coletta2022learning, coletta2021towards, li2020generating}, have shown promise in generating plausible market orders for simulations. However, GANs are susceptible to mode collapse \cite{thanh2020catastrophic} and training instability \cite{chu2020smoothness}, leading to a lack of realism and usefulness in the generated data. 

To address these shortcomings, we present our novel \textbf{TRA}nsformer-based Denoising \textbf{D}iffusion Probabilistic \textbf{E}ngine for LOB market \textbf{S}imulations (\textbf{TRADES}). TRADES generates realistic order flows that are time series, conditioned on past observations. We demonstrate that TRADES surpasses state-of-the-art (SoTA) methods in generating realistic and responsive market simulations. Importantly, due to its ability to handle multivariate time series generation, TRADES is adaptable to other domains requiring conditioned sequence data generation. Furthermore, TRADES readily adapts to an experimental agent introduced into the simulation, facilitating counterfactual market impact experiments.
In summary, our contributions are:
\begin{enumerate}
    \item \textbf{Realistic and responsive market simulation method:} We develop a Denoising Diffusion Probabilistic Engine for LOB Simulations (TRADES), exploiting a transformer-based neural network architecture. 
    \item \textbf{DeepMarket}: We release \textbf{DeepMarket}, the first open-source Python framework for market simulation with deep learning. 
    We also publish TRADES's implementation and checkpoints to promote reproducibility and facilitate comparisons and further research. 
    \item \textbf{Synthetic LOB dataset:} We release a synthetic LOB dataset composed of the TRADES's generated simulations. 
    We show in the results (section \ref{sec:results}) how the synthetic market data can be useful to train a deep learning model to predict stock prices. 
    \item \textbf{New ``world'' agent for market simulations: } We extend ABIDES \cite{byrd2020abides}, an agent-based simulation environment, in our framework by introducing a new world agent class accompanied by a simulation configuration, which, given in input a trained generative model, creates limit order book market simulations. Our experimental framework does not limit the simulation to a single-world agent but enables the introduction of other trading agents, which interact among themselves and with the world agent. This defines a hybrid approach between deep learning-based and agent-based simulations.
    \item \textbf{First quantitative metric for market simulations:} 
    The literature shows a notable absence of quantitative metrics to evaluate the generated market simulations. Typically, the evaluation relies on plots and figures. We posit that a robust and quantitative assessment is essential for the comparative analysis of various methodologies. To this end, we adapt the predictive score introduced in \cite{yoon2019time} to objectively and quantitatively evaluate the usefulness of the generated market data. 
    \item \textbf{Extensive experiments assessing usefulness, realism, and responsiveness:} We perform a suite of experiments to demonstrate that TRADES-generated market simulations abide by these three principles. 
    We show how TRADES outperforms SoTA methods \cite{byrd2020abides,coletta2022learning, coletta2021towards,vyetrenko2020get} according to the adopted predictive score and illustrate how TRADES follows the established stylized facts in financial markets \cite{vyetrenko2020get}.
\end{enumerate}

\section{Background}
Here, we provide background information on multivariate time series generation and limit order book markets. 
Furthermore, since TRADES is an extension of the Denoising Diffusion Probabilistic Model (DDPM), we summarize it in the Appendix \cite{bertiecaiappendix}.

\subsection{Multivariate time series generation}\label{sec:multivariate_ts}

Generating realistic market order streams can be formalized as a multivariate time series generation problem. Let $\mathbf{X} = \{\mathbf{x}_{1:N,1:K}\} \in \mathbb{R}^{N \times K}$, be a multivariate time series, where $N$ is the time dimension (i.e., length) and $K$ is the number of features. The goal of generating multivariate time series is to consider a specific time frame of previous observations, i.e., $\{\mathbf{x}_{1:N,1:K}\}$, and to produce the next sample $\mathbf{x}_{N+1}$. This task can easily be formulated as a self-supervised learning problem, where we leverage the past generated samples as the conditioning variable for an autoregressive model. In light of this, we can define the joint probability of the time series as in Eq.~\eqref{eq:joint_probability_ts}.
\begin{equation}\label{eq:joint_probability_ts}
    q(\mathbf{x}) = \prod_{n=1}^N q(\mathbf{x}_n \:|\: \mathbf{x}_1, ...,\, \mathbf{x}_{n-1})
\end{equation}
We leverage this concept at inference time using a sliding window approach. Hence, for every generation step,\footnote{For every $T$ diffusion steps there is a generation step} we generate a single sample $\mathbf{x}_N \in \mathbb{R}^K$. In the next step, we append the generated $\mathbf{x}_N$ to the end of the conditional part and shift the entire time series one step forward (see Section \ref{sec:4} for details with TRADES). Because we aim to generate a multivariate time series starting from observed values, we model the conditioned data distribution $q(\mathbf{x}_N | \mathbf{x}_{1:N-1})$ with a learned distribution $p_\theta(\mathbf{x}_N | \mathbf{x}_{1:N-1})$, to sample from it. Hereafter, we denote the conditional part\footnote{We remark that the conditioning can be composed of real or generated samples.} with $\mathbf{x}^c$, and the upcoming generation part with $\mathbf{x}^g$.

\subsection{Limit Order Book}
\label{sec:lob_description}
In a Limit Order Book (LOB) market, traders can submit orders to buy or sell a certain quantity of an asset at a particular price. There are three main types of orders in a limit order market. (1) A \textbf{market order} is filled instantly at the best available price. (2) A \textbf{limit order} allows traders to decide the maximum (in the case of a buy) or the minimum (in the case of a sell) price at which they want to complete the transaction. A quantity is always associated with the price for both types of orders. 
(3) A \textbf{cancel order}\footnote{Sometimes, events referring to these orders are defined as deletion.} removes an active limit order.
The Limit Order Book (LOB) is a data structure that stores and matches the active limit orders according to a set of rules. The LOB is accessible to all the market agents and is updated with each event, such as order insertion, modification, cancellation, and execution.
The most used mechanism for matching orders is the Continuous Double Auction (CDA). In a CDA, orders are executed whenever a price overlaps between the best bid (the highest price a buyer is willing to pay) and the best ask (the lowest price a seller is willing to accept).
This mechanism allows traders to trade continuously and competitively. 
The evolution over time of a LOB represents a multivariate temporal problem.
We can classify the research on LOB data into four main types of studies, namely empirical studies analyzing the LOB dynamics \cite{bouchaud2002statistical, cont2001empirical}, price and volatility forecasting \cite{zhang2019deeplob, sirignano2019deep},  modeling the LOB dynamics \cite{cont2011statistical, gould2013limit} and LOB market simulation \cite{byrd2020abides, coletta2021towards, li2020generating}.

\begin{figure*}[!t]
    \centering
    \includegraphics[width=\textwidth]{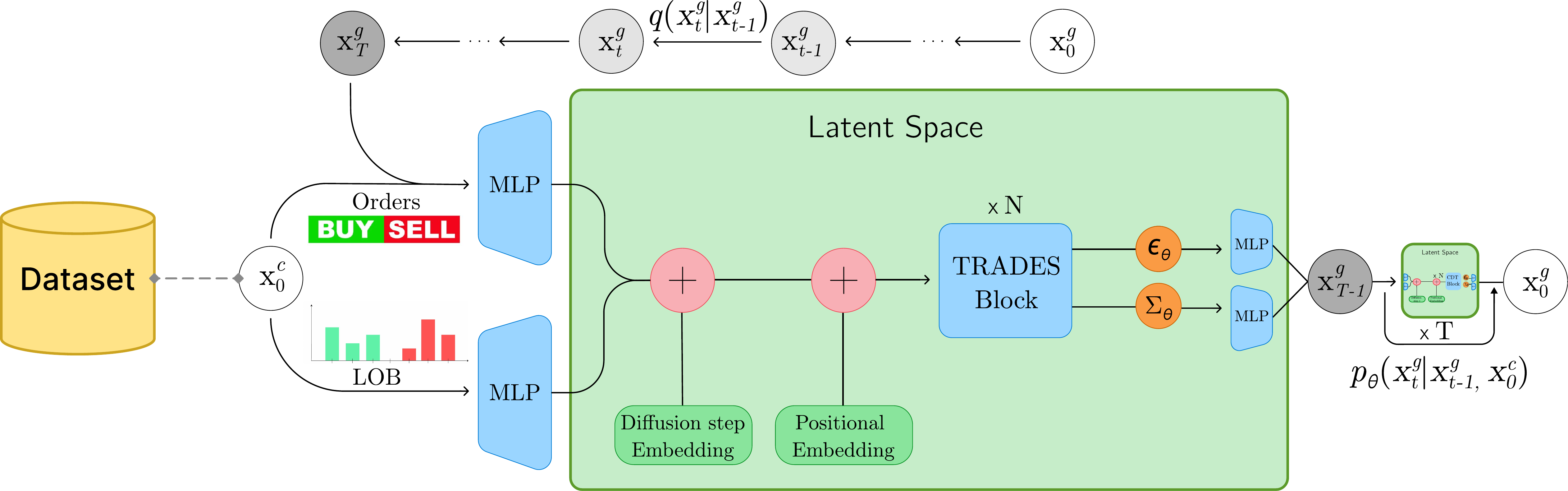}
    \caption{The training procedure and architecture of TRADES. We apply noise to the time series's last element until obtaining $x_T^g$. We condition $x_T^g$ on the previous $N-1$ orders and the last $N$ LOB snapshots. We feed these new tensors to two separate MLPs, each composed of two fully connected layers, to augment them into a higher-dimensional space. We concatenate the augmented output vectors on the features axis and sum the diffusion embedding step $t$ and the positional embedding; We feed the result to the TRADES modules, composed of multi-head self-attention and feedforward layers, producing the noise $\varepsilon_\theta$ and the standard deviation $\Sigma_\theta$. $\varepsilon_\theta$ and $\Sigma_\theta$ go through a de-augmentation phase via MLPs to map them back to the input space and are used to reconstruct $\mathbf{x}_{T-1}^g$. We repeat this procedure $T$ times until we recover the original $\mathbf{x}_0^g$.}
    \label{fig:architecture}
\end{figure*}
\section{Related Works}
\textbf{Diffusion models for time series generation.}
Diffusion models have been successfully applied to generate images \cite{austin2021structured}, video \cite{mei2023vidm}, and text \cite{zhu2023conditional}. Recently, they have also been exploited for time series forecasting \cite{li2023generative}, imputation \cite{tashiro2021csdi}, and generation \cite{lim2023regular}.
 Lim et al. \cite{lim2023regular} tackle time series generation using diffusion models. They present TSGM, which relies on an RNN-based encoder-decoder architecture with a conditional score-matching network. Yangming Li’s TS-Diffusion \cite{yuan2024diffusionts} introduces a neural-ODE encoder with self-attention and a diffusion core to manage irregularities and missingness in complex series

\noindent\textbf{Market simulation with deep learning.}
Generating realistic market simulations using deep learning is a new paradigm. Traditional computational statistical approaches \cite{cont2001empirical} and IABS \cite{byrd2020abides} rely on strong assumptions, such as constant order size, failing to produce realistic simulations. These methods are mainly used to study how 
the interactions of autonomous agents give rise to aggregate statistics and emergent phenomena in a system. Limit order book simulations are increasingly relying on deep learning. Li et al. \cite{li2020generating} were the first to leverage a Wasserstein GAN (WGAN) \cite{arjovsky2017wasserstein} for generating order flows based on historical market data. Similarly, Coletta et al.  \cite{coletta2021towards, coletta2022learning} employ WGANs in their stock market simulations, addressing the issue of responsiveness to experimental agents for the first time. Different
\\
\\
from \cite{coletta2021towards, coletta2022learning, li2020generating}, we condition with both the last orders and LOB snapshots, pushing the generation process toward a more realistic market simulation. Hultin et al. \cite{hultin2023generative} extend \cite{cont2010stochastic} and model individual features with separate conditional probabilities using RNNs. Instead of relying on GANs, which are prone to model collapse and 
instability \cite{chu2020smoothness}, and RNNs, often hampered by the vanishing gradient phenomenon, we exploit diffusion-based models with an underlying transformer architecture. 
Nagy et al. \cite{nagy2023generative} rely on simplified state-space models to learn long-range dependencies, tackling the problem via a masked language modeling approach. Shi and Cartlidge \cite{shi2023neural} introduce NS-LOB, a novel hybrid approach that combines a pre-trained neural Hawkes \cite{shi2022state} process with a multi-agent trader simulation. We refer the reader to \cite{jain2024limit} for a comprehensive review of limit order book simulations.
\section{Transformer-based Denoising Diffusion Probabilistic Engine for LOB Simulations}

\label{sec:4}
We introduce TRADES, a Transformer-based Denoising Diffusion Probabilistic Engine for LOB Simulations. Conditional diffusion models are better suited than standard diffusion models in generative sequential tasks because they can incorporate information from past observations that guide the generation process towards more specific and desired outputs. 
We formalize the reverse process for TRADES and the self-supervised training procedure. In Section \ref{sec:cdt_market_simulation}, we specialize our architecture for market simulations.

\subsection{Generation with TRADES}\label{sec:reverse_TRADES}
Here, we focus on an abstract time series generation task with TRADES. The goal of probabilistic generation is to approximate the true conditional data distribution $q(\mathbf{x}_0^g \, |\, \mathbf{x}_0^c)$ with a model distribution $p_{\theta}(\mathbf{x}_0^g \, |\, \mathbf{x}_0^c)$. During the forward process, we apply noise 
only to the ``future'' -- i.e., the part of the input we want to generate -- while keeping the observed values unchanged. Therefore, the forward process is defined as in the unconditional case (see Appendix \cite{bertiecaiappendix} or \cite{ho2020denoising}).
For the reverse process, we extend the unconditional 
\\
\\
\\
\\
one $p_{\theta}(\mathbf{x}_{0:T})$, to the conditional case:
\begin{equation}
p_{\theta}(\mathbf{x}_{0:T}^g) := p(\mathbf{x}_T^g) \prod_{t=1}^{T} p_{\theta} (\mathbf{x}_{t-1}^g \, |\, \mathbf{x}_t^g, \mathbf{x}^c_0)
\end{equation}
\begin{equation}
    p_{\theta}(\mathbf{x}_{t-1}^g \, |\, \mathbf{x}_t^g, \mathbf{x}^c_0) := \mathbf{\mathcal{N}} (\mathbf{x}_{t-1}^g; \: \boldsymbol{\mu}_\theta (\mathbf{x}_t^g, \mathbf{x}^c_0, t), \mathbf{\Sigma}_\theta (\mathbf{x}_t^g, \mathbf{x}^c_0, t))
\end{equation}
 We define the conditional denoising learnable function as in Eq.~\eqref{eq:cond_denoise_funct}.
\begin{equation}\label{eq:cond_denoise_funct}
    \boldsymbol{\epsilon}_{\theta} : \left( \mathbf{x}^g_{t} \in \mathbb{R}^{S\times K}, \: \mathbf{x}^c_0 \in \mathbb{R}^{M\times K}, \: t \in \mathbb{R}^{K} \right) \rightarrow \boldsymbol{\epsilon}_t \in \mathbb{R}^{S\times K}
\end{equation}%
where $M + S = N$. We set $M = N - 1$ and $S = 1$ for our experiments. Using $S > 1$ increases the efficiency, but also the task complexity because, in a single generative step, we generate $S$ time series steps. 
We exploit the parametrization proposed in \cite{ho2020denoising} described in Eq.~\eqref{eq:mean_parametrization} to estimate the mean term.
\begin{align}
\label{eq:mean_parametrization}
\boldsymbol{\mu}_\theta(\mathbf{x}^g_t, \mathbf{x}^c_0, t) = \frac{1}{\sqrt{\alpha_t}} \left(\mathbf{x}^g_t - \frac{\beta_t}{\sqrt{1 - \overbar{\alpha}_t}}\boldsymbol{\epsilon}_\theta(\mathbf{x}^g_t, \mathbf{x}^c_0, t)\right) \;\: \\ 
\text{where} \;\;\:  \mathbf{x}_t^g = \sqrt{\overbar{\alpha}_t}\mathbf{x}_0^g \, +\, \sqrt{1 - \overbar{\alpha}_t}\boldsymbol{\epsilon} \;\;\; \boldsymbol{\epsilon} \sim \mathcal{N}(\mathbf{0}, \mathbf{I}).
\end{align}
We do not rely on a fixed schedule as in \cite{ho2020denoising} regarding the variance term. Inspired by \cite{nichol2021improved}, we learn it as in Eq.~\eqref{eq:var_learning}.
\begin{equation}\label{eq:var_learning}
\mathbf{\Sigma}_\theta (\mathbf{x}_t^g, \mathbf{x}_0^c, t) = \exp({v \log \beta_t  +  (1 - v) \log \tilde{\beta}_t}),
\end{equation}
where $v$ is the neural network output, together with $\epsilon_{\theta}$. Nichol and Dhariwal \cite{nichol2021improved} found that this choice improves the negative log-likelihood, which we try to minimize.
After computing $\boldsymbol{\epsilon}_t$ and $\boldsymbol{\sigma}_t$, we denoise $\mathbf{x}_{t-1}^g$ as in Eq~\eqref{eq:denoise}.
\begin{equation}\label{eq:denoise}
    \mathbf{x}^g_{t-1} = \frac{1}{\sqrt{\alpha_t}} \left(\mathbf{x}^g_t - \frac{\beta_t}{\sqrt{1 - \overbar{\alpha}_t}}\boldsymbol{\epsilon}_\theta(\mathbf{x}^g_t, \mathbf{x}^c_0, t)\right) + \boldsymbol{\sigma}_t\mathbf{z}  
\end{equation}
where $\mathbf{z} \sim \mathcal{N}(\mathbf{0}, \mathbf{I})$. After the $T$ steps, the denoising process is finished, and $\mathbf{x}_0^g$ is reconstructed. 

\subsection{Self-supervised Training of TRADES}\label{sec:ss_TRADES}
\label{sec:train}
Given generation target $\mathbf{x}^g_0$ and conditional observations $\mathbf{x}^c_0$, we sample $\mathbf{x}_t^g = \sqrt{\overbar{\alpha}_t}\mathbf{x}_0^g \, +\, \sqrt{1 - \overbar{\alpha}_t}\boldsymbol{\epsilon}$, where $\boldsymbol{\epsilon} \sim \mathcal{N}(\mathbf{0}, \mathbf{I})$, and train $\boldsymbol{\epsilon}_{\theta}$ by minimizing Eq.~\eqref{eq:l_simple}.
\begin{equation}\label{eq:l_simple}
    \mathcal{L}_{\boldsymbol{\epsilon}}(\theta) := \mathbb{E}_{t, \mathbf{x}_0, \boldsymbol{\epsilon}} \Bigr[ ||\boldsymbol{\epsilon} - \boldsymbol{\epsilon}_\theta( \mathbf{x}_t^g, \mathbf{x}_0^c, t) ||^2 \Bigr]. 
\end{equation}
Inspired by \cite{nichol2021improved}, we  also learn\footnote{Notice from Eq~\eqref{eq_app:6} of the original DDPM formulation that $\mathbf{\Sigma}_\theta$ is fixed.} $\mathbf{\Sigma}_\theta$ optimizing it according to Eq.~\eqref{eq:vlb}.
\begin{equation}
\label{eq:vlb}
\begin{gathered}
 \mathcal{L}_{\mathbf{\Sigma}}(\theta):= \mathbb{E}_{q} \Biggr[\underbrace{-p_{\theta}(\mathbf{x}_0^g \, | \,  \mathbf{x}_1^g, \mathbf{x}_0^c)}_\text{$L_0$} + \underbrace{D_{KL}(q(\mathbf{x}_{T}^g \, | \, \mathbf{x}_0^g)\, ||\, p_{\theta}(\mathbf{x}_{T}^g)}_\text{$L_T$} \\+ \sum_{t=2}^T\underbrace{D_{KL}(q(\mathbf{x}_{t-1}^g \, | \, \mathbf{x}_t^g, \mathbf{x}_0^g) \, || \, p_{\theta}(\mathbf{x}^g_{t-1} \, | \, \mathbf{x}^g_t, \mathbf{x}_0^c))}_\text{$L_{t-1}$} \Biggr].
\end{gathered}
\end{equation}
We optimize Eq.~\eqref{eq:vlb} to reduce the negative log-likelihood, especially during the first diffusion steps where $\mathcal{L}_{\Sigma}$ is high \cite{nichol2021improved}. The final loss function is a linear combination of the two as in Eq:~\eqref{eq:final_loss}.
\begin{equation}\label{eq:final_loss}
    \mathcal{L} = \mathcal{L}_{\boldsymbol{\epsilon}} + \lambda\mathcal{L}_{\mathbf{\Sigma}}.
\end{equation}
Where $\lambda$ is used to balance the two terms. We perform training by relying on a self-supervised approach as follows.  Given a time series $\mathbf{x}_0 \in \mathbb{R}^{N \times K}$, we apply noise only to its last element -- i.e., $\mathbf{x}^N_0$ -- through the forward pass. Then, we denoise it via the reverse process and learn $p_{\theta}(\mathbf{x}^g_{t-1} \, | \, \mathbf{x}^g_t, \mathbf{x}_0^c)$ which aims to generate a new sample from the last observations. Therefore, during training, the conditioning has only observed values.
At sampling time, TRADES generates new samples autoregressively, conditioned on its
previous outputs, until the simulation ends.

\section{TRADES for Market Simulation}\label{sec:cdt_market_simulation}

To create realistic and responsive market simulations, we implement TRADES to generate orders conditioned on the market state. TRADES's objective is to learn to model the distribution of orders. Fig. \ref{fig:architecture} presents an overview of the diffusion process and the architecture. The network that produces $\varepsilon_\theta$ and $\Sigma_\theta$ contains several transformer encoder layers to model the temporal and spatial relationships of the financial time series \cite{sirignano2019deep}.
Since transformers perform better with large dimension tensors, we project the orders tensor and the LOB snapshots 
to a higher dimensional space, using two fully connected layers. Hence, TRADES operates on the augmented vector space. After the reverse process, we de-augment $\varepsilon_\theta$ and $\Sigma_\theta$ projecting them back to the input space to reconstruct $\mathbf{x}_{t-1}^g$ and compute the loss. 

\noindent\textbf{Conditioning.}
The conditioned diffusion probabilistic model learns the conditioned probability distribution $p_{\theta}(o | s)$, where $s$ is the market state, and $o$ is the newly generated order, represented as $(p, q, d, b, \delta, c)$, where $p$ is the price, $q$ is the quantity, $d$ is the direction either sell or buy, $b$ is the depth, i.e., the difference between $p$ and the best available price, $\delta$ is the time offset representing the temporal distance from the previously generated order and $c$ is the order type -- either market, limit or cancel order. The asymmetries between the buying and selling side of the book indicate shifts in the supply and demand curves caused by exogenous and unobservable factors that influence the price \cite{cao2009information}. Therefore, as depicted in Fig. \ref{fig:architecture}, the model's conditioning extends beyond the last $N-1$ orders. To effectively capture market supply and demand dynamics encoded within the LOB, we incorporate the last $N$ LOB snapshots of the first $L$ LOB levels as input, where each level has a bid price, bid size, ask price, and ask size. We set $N = 256$ as in \cite{coletta2021towards}, and $L = 10$. 
We argue that this choice of $L$ is a reasonable trade-off between conditioning complexity and feature informativeness. 
Several works \cite{huang1994market, tran2022informative, pascual2003pieces, cao2008order, cao2009information} have shown that the orders behind the best bid and ask prices play a significant role in price discovery and reveal information about short-term future price movements. 
In Sec. \ref{sec:ablation}, we delve deeper into the conditioning choice and method, performing an ablation and a sensitivity study.

\section{DeepMarket framework with synthetic dataset for deep learning market simulations}
We present DeepMarket, an open-source Python framework developed for LOB market simulation with deep learning. DeepMarket offers the following features: (1) pre-processing for high-frequency market data; (2) a training environment implemented with PyTorch Lightning; (3) hyperparameter search facilitated with WANDB; (4) TRADES and CGAN implementations and checkpoints to directly generate a market simulation without training; (5) a comprehensive qualitative (via the plots in this paper) and quantitative (via the predictive score) evaluation. 
To perform the simulation with our world agent and historical data, we extend ABIDES \cite{byrd2020abides}, an open-source agent-based interactive Python tool. 
\subsection{TRADES-LOB: a new synthetic LOB dataset}
In LOB research, one major problem is the unavailability of a large LOB dataset. In fact, if you want to access a large LOB dataset, you need to pay large fees to a data provider. 
The only two freely available LOB datasets are \cite{huang2021benchmark} and \cite{urn:nbn:fi:csc-kata20170601153214969115}, which have a lot of limitations.
The high cost and low availability of LOB data restrict the application and development of deep learning algorithms in the LOB research community.
In order to foster collaboration and help the research community, we release a synthetic LOB dataset: TRADES-LOB. TRADES-LOB comprises simulated TRADES market data for Tesla and Intel for two days. Specifically, the dataset is structured into four CSV files, each containing 50 columns. The initial six columns delineate the order features, followed by 40 columns that represent a snapshot of the LOB across the top 10 levels. The concluding four columns provide key financial metrics: mid-price, spread, order volume imbalance, and Volume-Weighted Average Price (VWAP), which can be useful for downstream financial tasks, such as stock price prediction. In total, the dataset is composed of 265,986 rows and 13,299,300 cells, which is similar in size to the benchmark FI-2010 dataset \cite{urn:nbn:fi:csc-kata20170601153214969115}. 
The dataset will be released with the code in the GitHub repository. 
We show in the results (section \ref{sec:results}) how the synthetic market data can be useful to train a deep learning model to predict stock prices.
\section{Experiments}\label{sec:experiments}
\begin{figure*}[!t]
    \centering
    \begin{subfigure}{0.32\linewidth}
    \includegraphics[width=\textwidth]{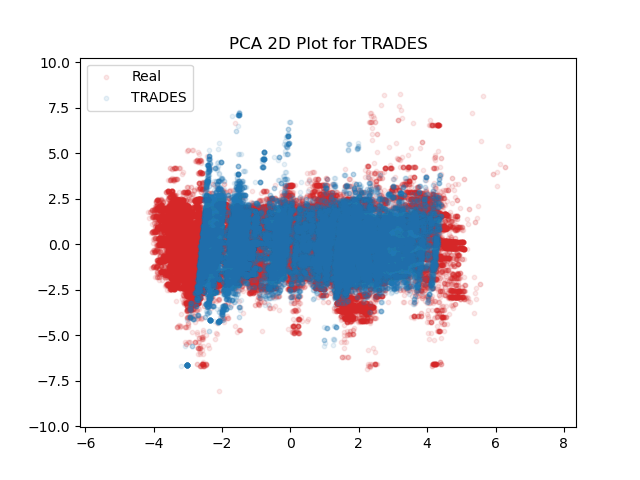}
    \end{subfigure}
    \begin{subfigure}{0.32\linewidth}
        \includegraphics[width=\textwidth]{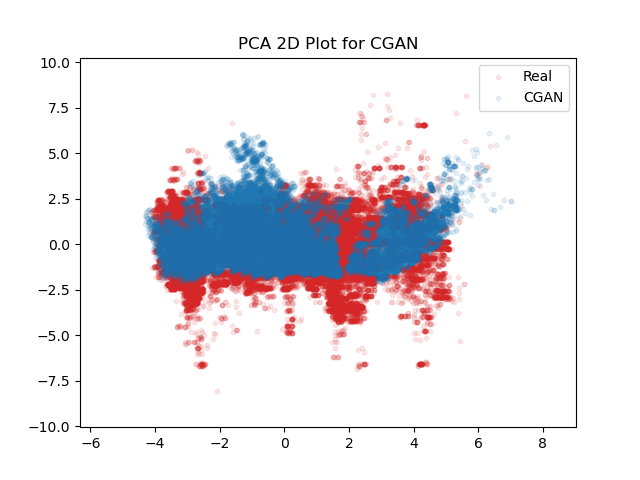}
    \end{subfigure}
    \begin{subfigure}{0.32\linewidth}
        \includegraphics[width=\textwidth]{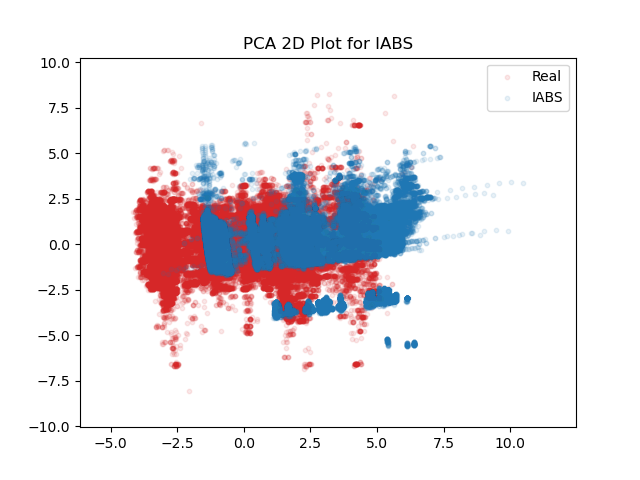}
    \end{subfigure}
    
    \caption{PCA analysis on TSLA 29/01. TRADES covers 67.04\% of the real distribution, better than the other two methods (52.92\% for IABS and 57.49\% for CGAN).}
    \label{fig:PCA}

\end{figure*}

\noindent\textbf{Dataset and reproducibility.}
\label{dataset}
In almost all SoTA papers in this subfield, the authors use one, two, or three stocks \cite{shi2023neural, coletta2021towards, nagy2023generative, li2020generating, shi2022state, hultin2023generative}, most of which are tech. Following this practice, we create a LOB dataset from two NASDAQ stocks\footnote{The data we used are downloadable from \url{https://lobsterdata.com/tradesquotesandprices} upon buying the book indicated on the website.} -- i.e., Tesla and Intel -- from January 2nd to the 30th of 2015. We argue that stylized facts and market microstructure behaviors, which are the main learning objective of TRADES, are independent of single-stock behaviors 
(see \cite{bouchaud2009markets, bouchaud2002statistical, cont2014price, gould2013limit}), so the particular stock characteristics, such as volatility, market cap, and p/e ratio, are not fundamental. Each stock has 20 order books and 20 message files, one for each trading day per stock, totaling $\sim$24 million samples. The message files contain a comprehensive log of events from which we select market orders, limit orders, and cancel orders.
Each row of the order book file is a tuple $\big(P^{ask}(t), V^{ask}(t), P^{bid}(t), V^{bid}(t)\big)$ where $P^{ask}(t)$ and $P^{bid}(t) \in \mathbb{R}^L$ are the prices of levels $1$ through $L$, and $V^{ask}(t)$ and $V^{bid}(t) \in \mathbb{R}^L$ are the corresponding volumes. We use the first 17 days for training, the 18th day for validation, and the last 2 for market simulations. We are aware of the widely used FI-2010 benchmark LOB dataset \cite{urn:nbn:fi:csc-kata20170601153214969115} for stock price prediction. However, the absence of message files in this dataset hinders simulating the market since the orders cannot be reconstructed. In the Appendix \cite{bertiecaiappendix}, we provide an overview of FI-2010 and its limitations. 

\noindent\textbf{Experimental setting.} After training the model for $70,000$ steps until convergence, we freeze the layers and start the market simulation. A simulation is composed of (1) the electronic market exchange that handles incoming orders and transactions; (2) the TRADES-based ``world'' agent, which generates new orders conditioned on the market state; and (3) one or more optional experimental agents, that follow a user-customizable trading strategy, enabling counterfactual and market impact experiments. So, the experimental framework is a hybrid approach between a deep learning model and an interactive agent-based simulation.

We conduct the simulations with the first 15 minutes of real orders to compare the generated ones with the market replay.\footnote{Market replay denotes the simulation performed with the real historical orders of that day.} Afterward, the diffusion model takes full control and generates new orders autoregressively, conditioned on its previous outputs, until the simulation ends. 
After the world agent generates a new order, there is a post-processing phase in which the output is transformed into a valid order. We begin the simulation at 10:00 and terminate it at 12:00. This choice ensures that the generated orders are sufficient for a thorough evaluation while maintaining manageable processing times. On average, 50,000 orders are produced during this two-hour time frame. The output CSV file of the simulation contains the full list of orders and LOB snapshots of the simulation. 
All experiments are performed with an RTX 3090 and a portion of an A100.
In the Appendix \cite{bertiecaiappendix}, we detail the data pre- and post-processing and model hyperparameter choice. 

\noindent\textbf{Baselines.} We compare TRADES with the Market Replay -- i.e., ground truth (market replay) -- a IABS configuration, and the Wasserstein GAN -- i.e.,  CGAN -- under the setting of \cite{coletta2022learning}, similar to the same of those proposed in \cite{coletta2021towards, coletta2023conditional}. We implemented CGAN from scratch given that none of the implementations in \cite{coletta2023conditional, coletta2022learning, coletta2021towards} are available. We report details in the Appendix \cite{bertiecaiappendix}. Regarding IABS configuration, we used the Reference Market Simulation Configuration, introduced in \cite{byrd2020abides}, which is widely used as comparison \cite{coletta2021towards, coletta2022learning, vyetrenko2020get}. The configuration includes 5000 noise, 100 value, 25 momentum agents, and 1 market maker.\footnote{the full specifics are in the GitHub page of ABIDES in the tutorial section.} We do not compare with other SoTA methods due to the unavailability of open-source implementations and insufficient details to reproduce the results. In some cases, the code is provided but the results are not reproducible due to computational constraints.

\subsection{Results}
\label{sec:results}
Here, we evaluate the \textit{usefulness}, \textit{realism}, and the \textit{responsiveness} of the generated market simulations for Tesla and Intel. We train two TRADES Models, one for each stock. After training, we freeze the models and use them to generate orders during the simulation phase. A major disadvantage of market simulation is the misalignment in the literature for a single evaluation protocol. All the other works \cite{coletta2022learning, hultin2023generative,nagy2023generative, coletta2021towards} analyze performances with plots, hindering an objective comparison between the models. To fill this gap, we adapt the \textit{predictive score} \cite{yoon2019time} for market simulation. Predictive score is measured as an MAE, by training a stock mid-price predictive model on synthetic data and testing it on real data, that evaluates the usefulness of the generated simulations. In the Appendix \cite{bertiecaiappendix}, we detail the computation of the predictive score.
\begin{table}[h]
    \centering
    \caption{Average predictive score (MAE) over two days on Tesla and Intel stocks. Bold values show the best MAE.} 
    \label{tab:predictive_scores_cdt_vs_soa}
    \resizebox{.6\linewidth}{!}{%
        \begin{tabular}{c|cc}
            \toprule
            & \multicolumn{2}{c}{\textbf{Predictive Score}$\downarrow$} \\
            \cmidrule(lr){2-3} 
            \textbf{Method} & \textbf{Tesla} & \textbf{Intel} \\
            \midrule
            Market Replay & 0.923 & 0.149 \\
            \midrule
            IABS & 1.870 & 1.866 \\
            CGAN & 3.453 & 0.699 \\
            TRADES & \textbf{1.213} & \textbf{0.307} \\
            \bottomrule
        \end{tabular}
    }
\end{table}
\\
\noindent\textbf{Usefulness: TRADES outperforms the second-best by a factor of $\mathbf{\times3.27}$ and $\mathbf{\times 3.48}$ on both stocks.}\footnote{The values are computed dividing the second best value with the TRADES values, both subtracted by the market replay predictive score.}  
We report both stocks' average predictive scores in Table \ref{tab:predictive_scores_cdt_vs_soa}. Notice that the market replay scores represent the desired MAE of each model. The table reveals that TRADES exhibits performances approaching that of the real market replay, with an absolute difference of $0.29$ and $0.158$ from market replay, respectively, for Tesla and Intel, suggesting a diminishing gap between synthetic and real-data training efficacy. Interestingly, although IABS cannot capture the complexity of real trader agents, it outperforms CGAN on Tesla, while it remains the worst-performing strategy on Intel. 
In conclusion, we demonstrated how a predictive model trained with TRADES's generated market data can effectively forecast mid-price. 
\begin{figure}[!t]
    \centering
    \begin{subfigure}{0.45\linewidth}
    \includegraphics[width=\textwidth]{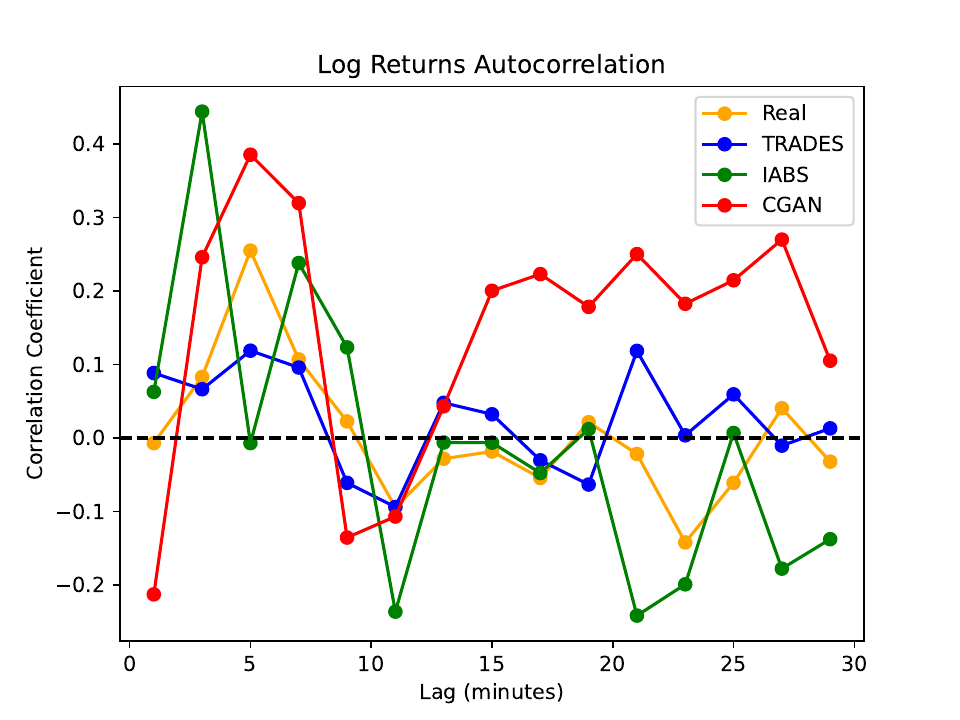}
    \end{subfigure}
    \begin{subfigure}{0.45\linewidth}
        \includegraphics[width=\textwidth]{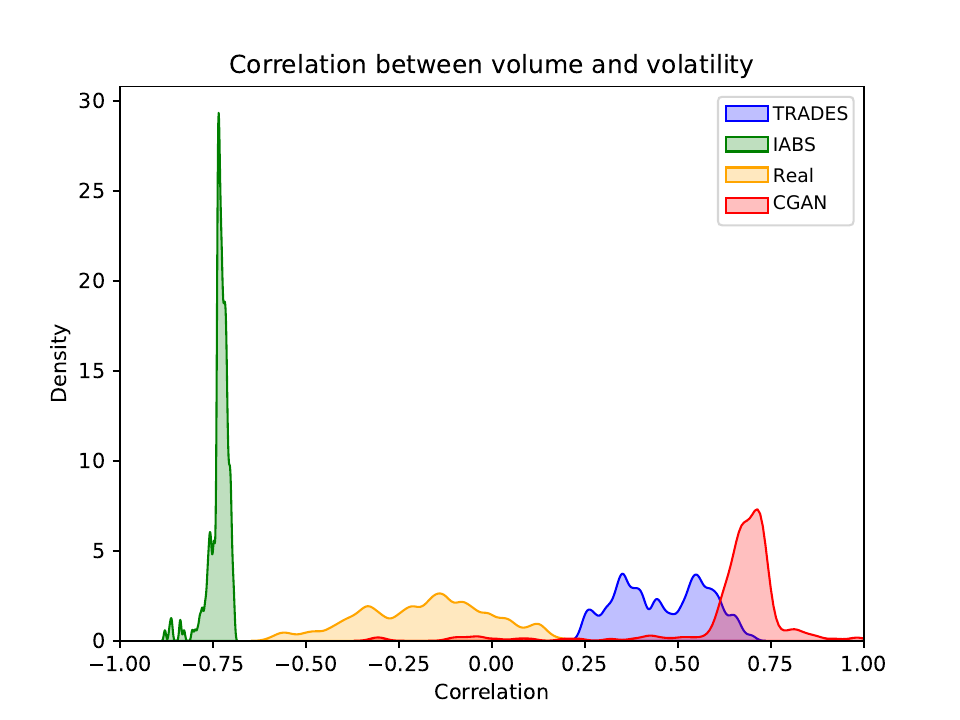}
    \end{subfigure}
    \begin{subfigure}{0.45\linewidth}
        \includegraphics[width=\textwidth]{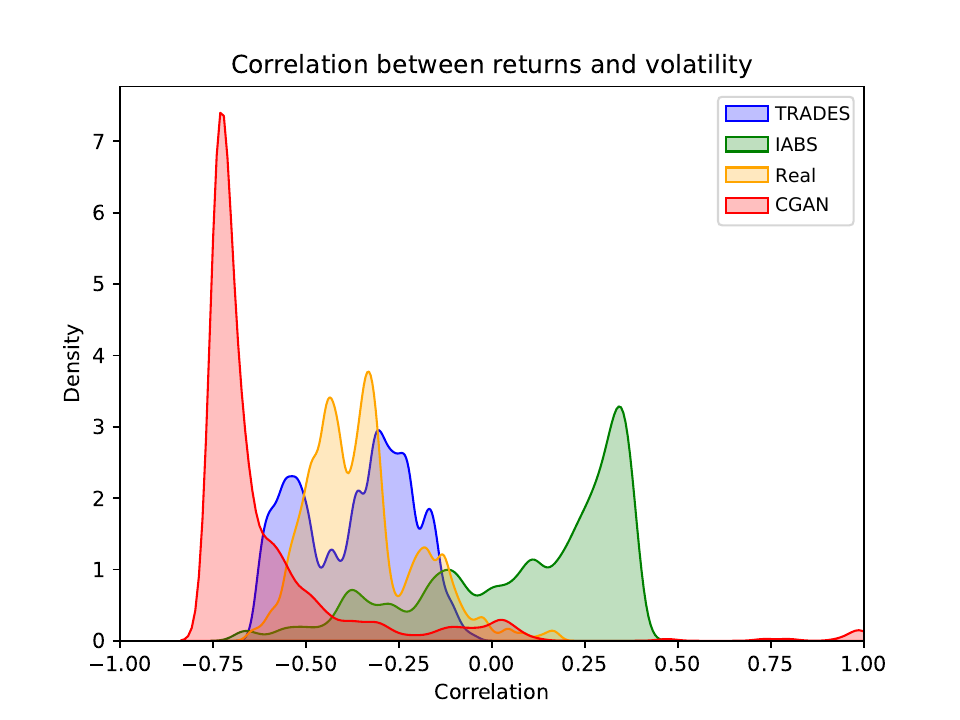}
    \end{subfigure}
    \begin{subfigure}{0.45\linewidth}
    \includegraphics[width=\textwidth]{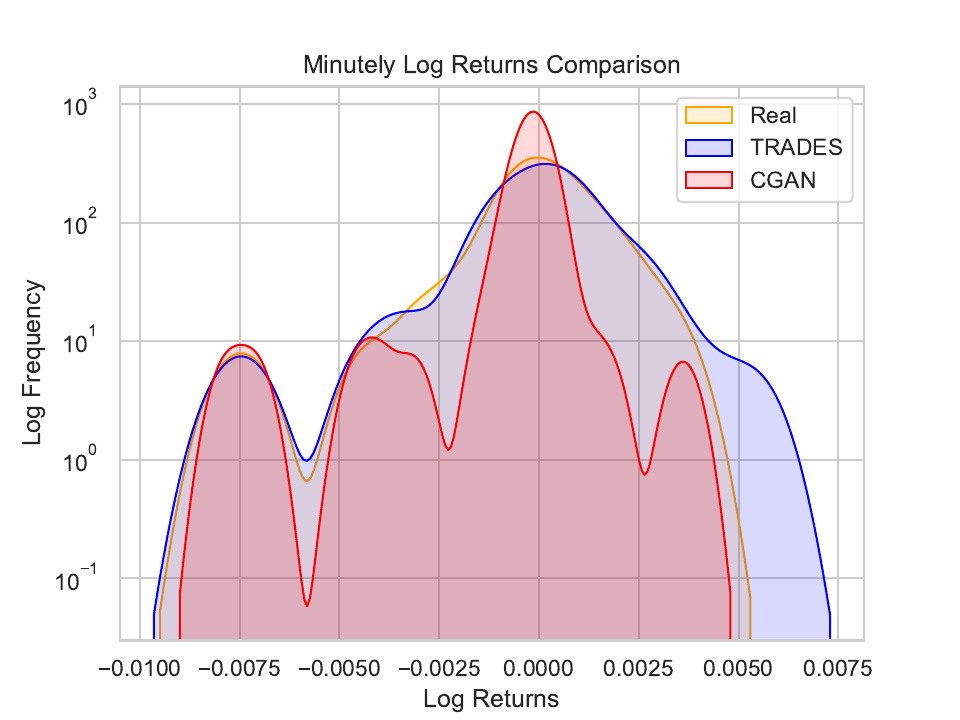}
    \end{subfigure}
    \begin{subfigure}{0.45\linewidth}
        \includegraphics[width=\textwidth]{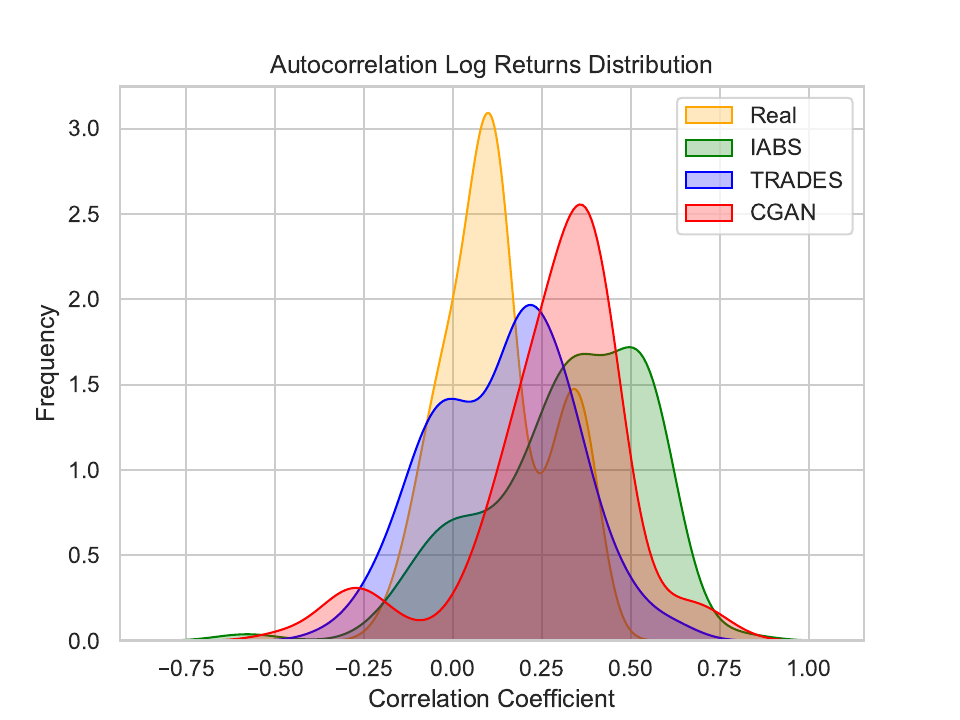}
    \end{subfigure}
    \begin{subfigure}{0.45\linewidth}
        \centering
        \includegraphics[width=\textwidth]{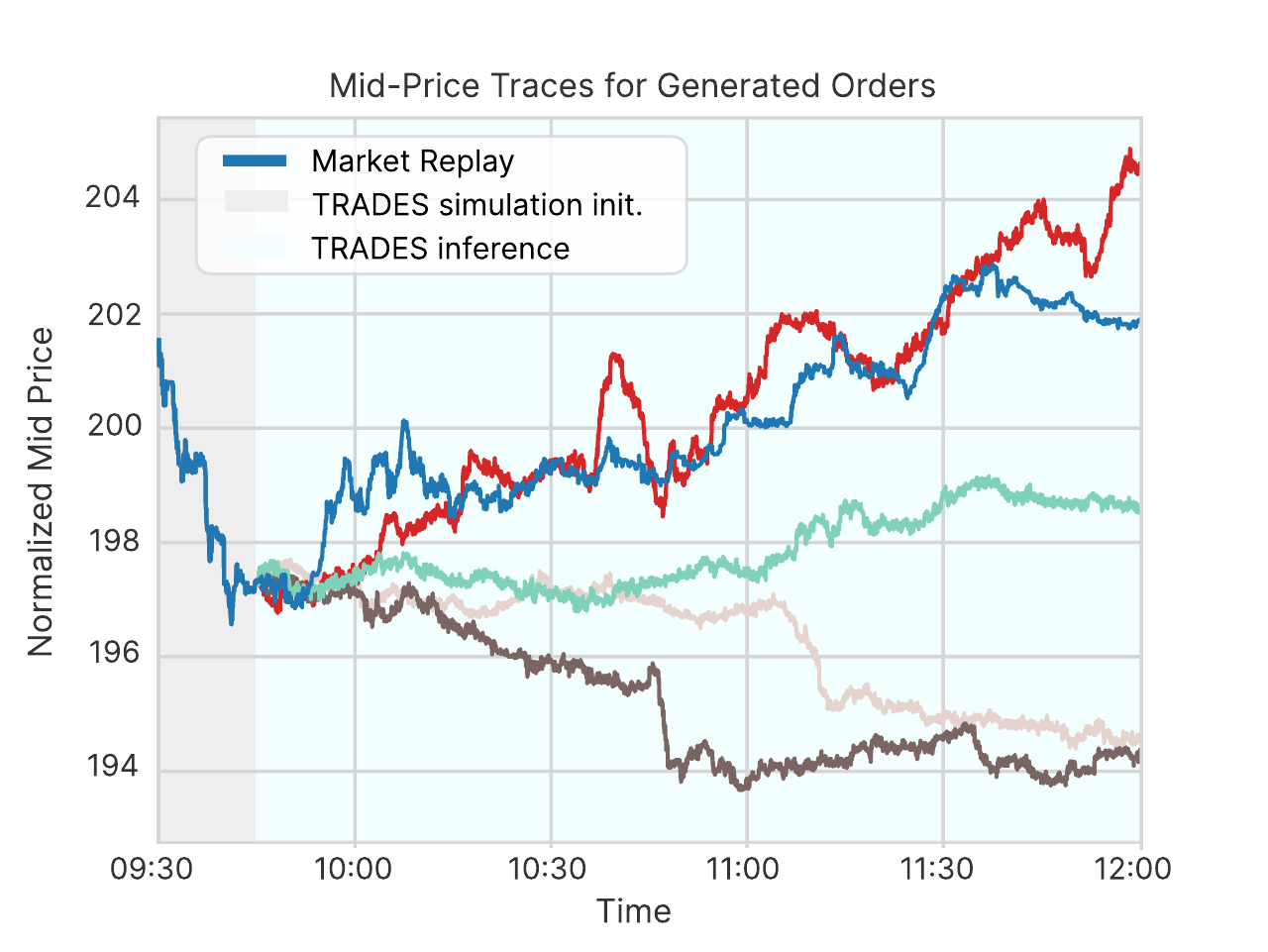}
    \end{subfigure}
    \caption{Stylized facts on Tesla 29/01. (1) Log returns autocorrelation. (2) The correlation between volume and volatility, and (3) between returns and volatility. (4) Comparison of the minute Log Returns distribution and (5) autocorrelation. (6) Mid-price traces of five different TRADES simulations.}
    \label{fig:stylized}
\vspace{0.3cm}
\end{figure}
\begin{figure}[!h]
    \centering
    \begin{subfigure}{0.45\linewidth}
    \includegraphics[width=\textwidth]{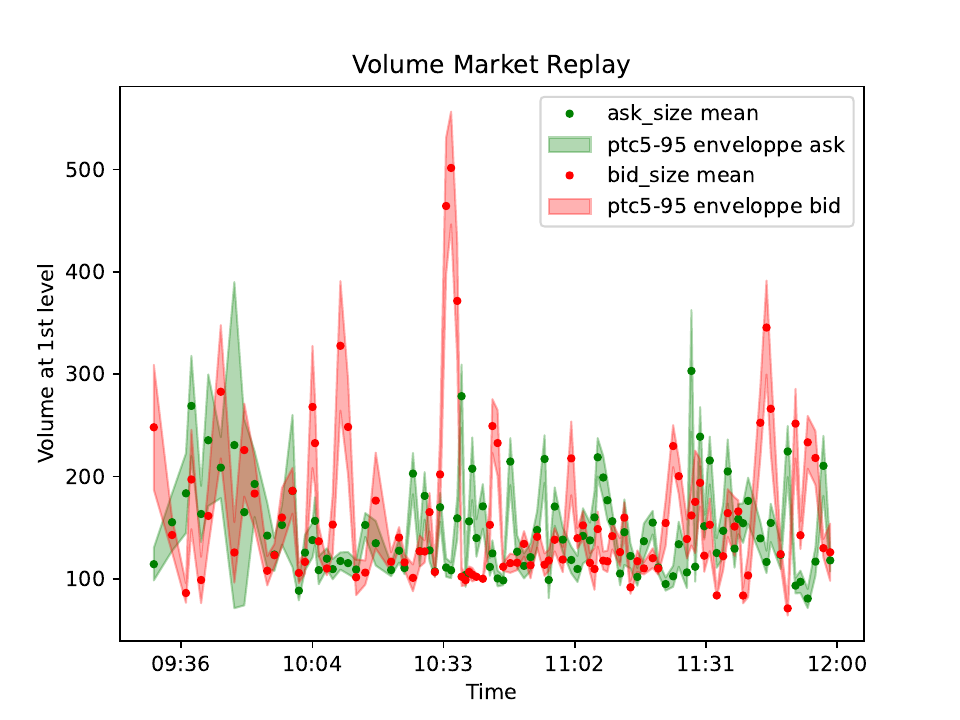}
    \end{subfigure}
    \begin{subfigure}{0.45\linewidth}
    \includegraphics[width=\textwidth]{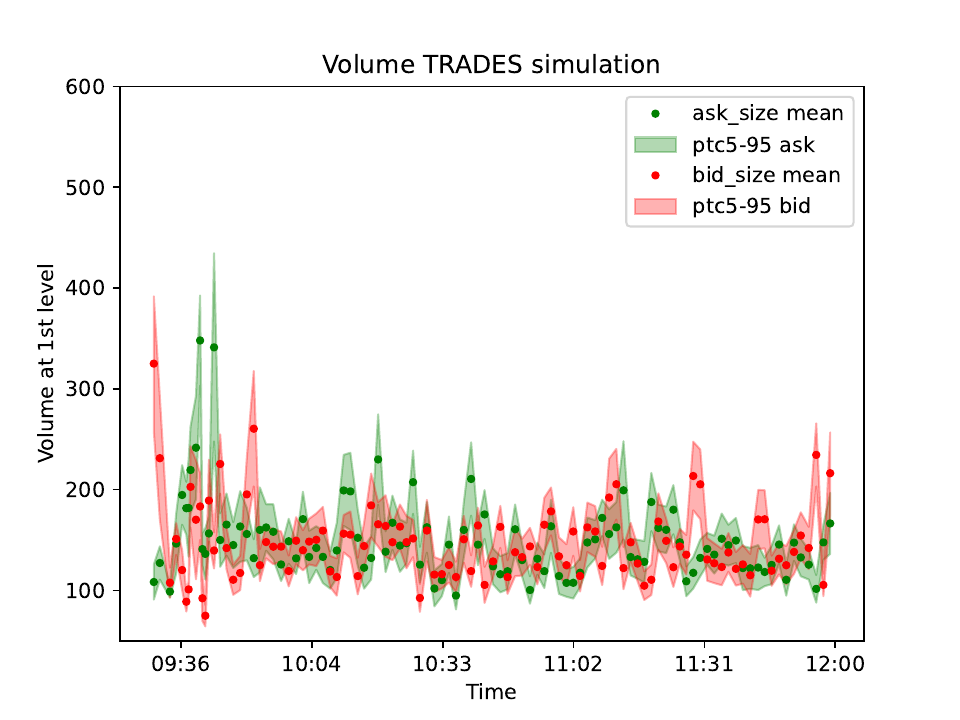}
    \end{subfigure}
    \begin{subfigure}{0.45\linewidth}
        \includegraphics[width=\textwidth]{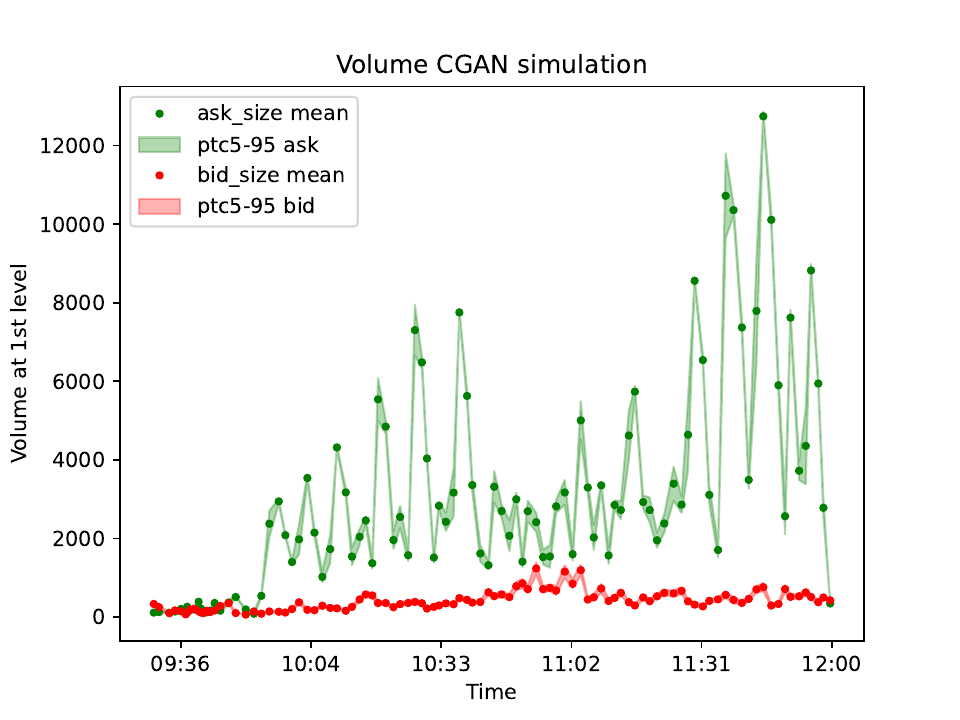}
    \end{subfigure}
    \begin{subfigure}{0.45\linewidth}
        \includegraphics[width=\textwidth]{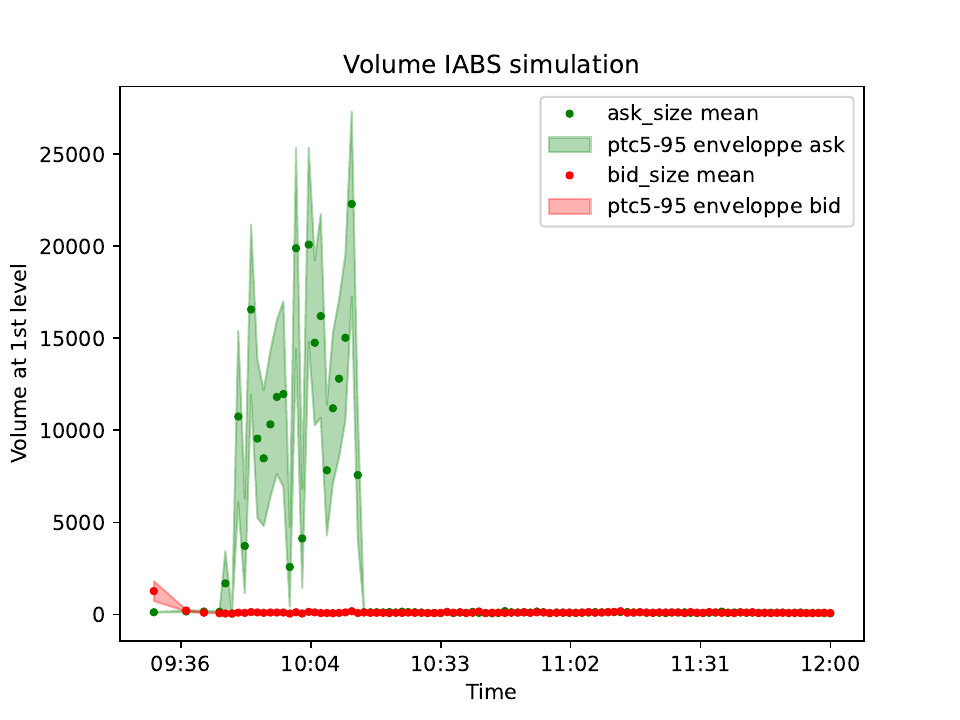}
    \end{subfigure}
    \caption{Volume at the first level of LOB on TSLA 29/01.}
    \label{fig:volume}
\end{figure}
\begin{figure*}[!h]
    \centering
    \begin{subfigure}{.45\linewidth}
    \includegraphics[width=\linewidth]{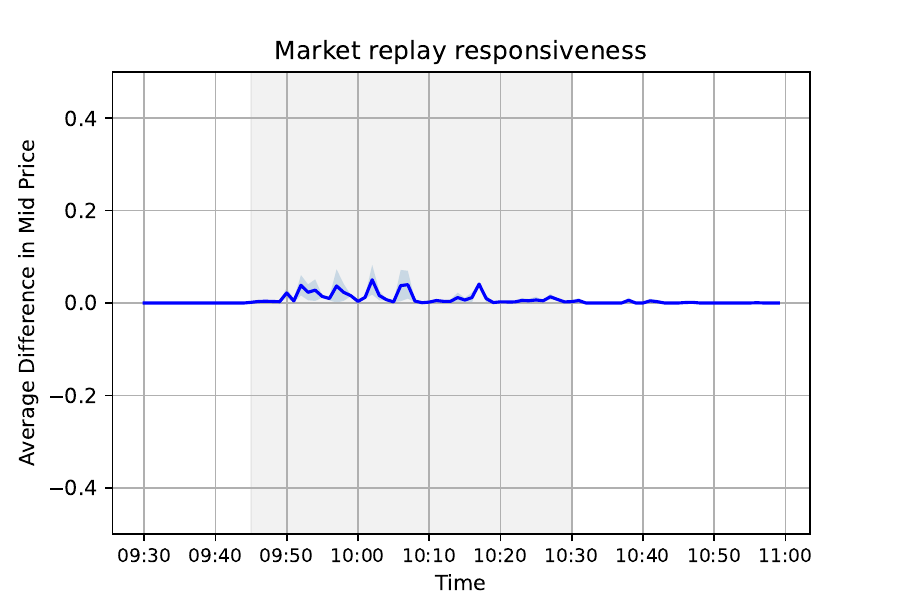}
    \end{subfigure}
    \begin{subfigure}{.45\linewidth}
    \includegraphics[width=\linewidth]{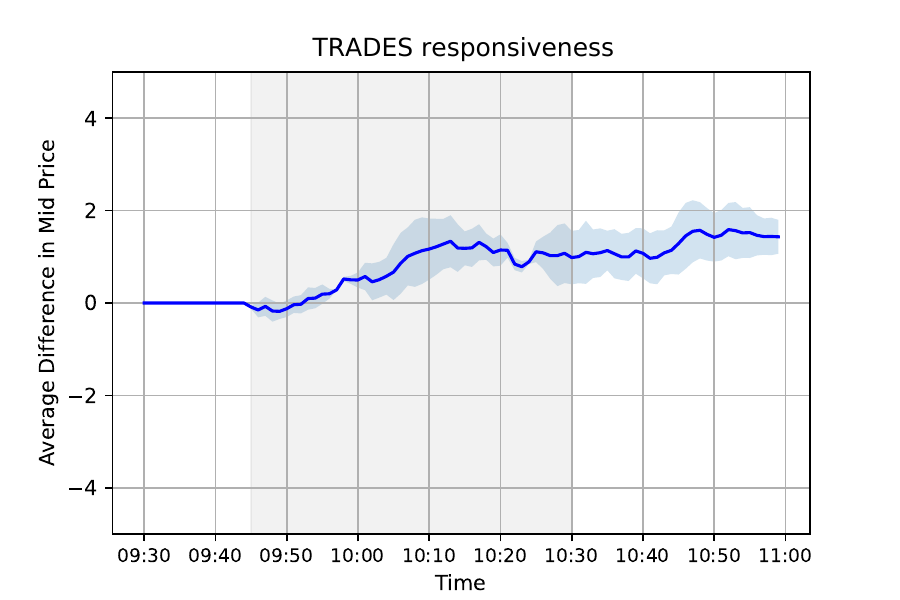}
    \end{subfigure}
    \caption{Average mid-price difference of market replay simulations and TRADES simulations with (shaded part) and without a POV agent (unshaded part), on 5 different seeds.}
        \label{fig:responsiveness}
\end{figure*}

\\
\noindent\textbf{Realism: TRADES covers the real data distribution and emulates many stylized facts.}
To evaluate the realism of the generated time series, we compare the real data distributions with those of the generated one. 
We employ a combination of Principal Component 
\\
\\
\\
Analysis (PCA), alongside specialized financial methodologies that include the comparison of stylized facts and important features.
In Fig. \ref{fig:PCA}, we show PCA plots to illustrate how well the synthetic distribution covers the original data. TRADES (blue points) cover 67.04\% of the real data distribution (red points) according to a Convex Hull intersection compared to 52.92\% and 57.49\% of CGAN and IABS, respectively.
Following \cite{vyetrenko2020get}, we evaluate the simulation realism by ensuring that the generated market data mimics stylized facts derived from real market scenarios. Fig. \ref{fig:stylized} (1) illustrates that TRADES, similarly to the market replay and differently to IABS and CGAN, obey the expected \textit{absence of autocorrelation}, indicating that the linear autocorrelation of asset returns decays rapidly, becoming statistically insignificant after 15 minutes. Fig. \ref{fig:stylized} (2) highlights TRADES's resemblance with \textit{the positive volume-volatility correlation}. 
TRADES's generated orders show a positive volume-volatility correlation.
Interestingly, TRADES captures this phenomenon better than that particular market replay day. 
Recall that TRADES is trained on 17 days of the market, so it correctly captures this correlation.
 We acknowledge that the real market might not always respect all the stylized facts due to their inherent non-deterministic and non-stationary nature. Also, CGAN resembles this phenomenon but, differently from TRADES, with an unrealistic intensity.

Fig \ref{fig:stylized} (3) shows that TRADES exhibits \textit{asset returns and volatility negative correlation}, emulating the market replay distribution in contrast with IABS. Similar to the previous case, also, CGAN also resembles this phenomenon but with an unrealistic intensity. Fig. \ref{fig:stylized} (4) illustrates that TRADES almost perfectly resembles the real distribution in terms of \textit{log returns}. We leave IABS out since it disrupts the plot's scale. In Fig. \ref{fig:stylized} (5), we show the autocorrelation function of the squared returns, commonly used as an analytical tool to measure the degree of \textit{volatility clustering}: i.e., high-volatility episodes exhibit a propensity to occur in close temporal proximity. Note that TRADES emulates the real distribution better than the other two methods.
In Fig. \ref{fig:stylized} (6), we illustrate the mid-price time series of five different TRADES\footnote{TRADES is frozen, and the same model is used for all simulations.} simulations on the same day of Tesla (29/01) and the market replay of that day. 
The mid-price traces generated show diversity and realism.
Lastly, to consolidate our claims about TRADES's realism, in Fig. \ref{fig:volume} we also analyze the volume distribution of the first LOB level. Notice how the scale and the overall behavior of the volume time series in the market replay strongly correlate with that of TRADES's simulation, while it is completely different w.r.t. SoTA approaches.

\noindent\textbf{Responsiveness: TRADES is responsive to external agent.} 
The responsiveness of a LOB market simulation generative model is crucial, especially if the objective of the market simulation is to verify the profitability of trading strategies or perform market impact experiments.
Generally \cite{coletta2021towards, coletta2022learning, shi2023neural}, the responsiveness of a generator is assessed through a market impact experiment (A/B test). Therefore, we conducted an experiment running some simulations w/ and w/o a Percentage-Of-Volume (POV) agent, which wakes up every minute and places a bunch of buy orders, until either $\phi$ shares have been transacted or the temporal window ends. We refer the reader to the Appendix \cite{bertiecaiappendix} for the details of the settings of this experiment. Fig. \ref{fig:responsiveness} depicts the normalized mid-price difference between the simulations w/ and w/o the POV agent for the market replay and TRADES.
Results are averaged over 5 runs. As expected, the historical market simulation exhibits only instantaneous impact \cite{gould2013limit}, that is the direct effect of the agent’s orders, which rapidly vanishes. Contrarily, the diffusion-based simulations demonstrate substantial deviation from the baseline simulation without the POV agent, altering the price permanently.
Quantifying the permanent price impact in real markets poses a significant challenge, as it requires comparing price differences between scenarios where a specific action took place and those where it did not. Such scenario analysis is not feasible with empirical data. However, by using TRADES-generated realistic simulations, this analysis becomes both feasible and measurable. In fact, the simulations allow us to run identical scenarios both with and without additional trader agents whose strategy can be fully defined by the user. 
In conclusion, the observed market impact in the TRADES simulations aligns with real market observations \cite{gould2013limit, bouchaud2009markets}, enabling the evaluation of trading strategies\footnote{we want to be clear that it technically enables evaluating trading strategies, but it does not assure any profitability in a real market scenario.} and counterfactual experiments.

\subsection{DDIM sampling}
One of the known limitations of diffusion models is the sampling time.
Indeed, the generation of a single sample necessitates hundreds of iterative passes through the neural network. 
In this work, the model was trained using a diffusion process comprising 100 steps. 
Recently, the Denoising Diffusion Implicit Model (DDIM) sampling method was proposed in \cite{song2020denoising} to speed up the generative process. 
Given that each hour of market simulation required six hours of computation on an RTX 3090, accelerating the simulation process was a relevant improvement.
Consequently, we conducted simulations employing 
DDIM sampling ($\eta = 0$), which is deterministic, utilizing a single step for each order.
We use the same trained model.
The results, presented in Table \ref{tab:ddim}, demonstrate that the performance degradation is significant but not disastrous despite a remarkable 100-fold increase in computational efficiency.
\begin{table}[h]
    \centering
    \caption{Average predictive score (MAE) over two days on Tesla and Intel stocks. DDIM sampling is done with a single step, while DDPM with 100.} 
    \label{tab:ddim}
    \resizebox{.6\linewidth}{!}{%
        \begin{tabular}{c|cc}
            \toprule
            & \multicolumn{2}{c}{\textbf{Predictive Score}$\downarrow$} \\
            \cmidrule(lr){2-3} 
            \textbf{Method} & \textbf{Tesla} & \textbf{Intel} \\
            \midrule
            DDIM & 3.146 & 0.486 \\
            DDPM (orig.) & \textbf{1.213} & \textbf{0.307} \\
            \bottomrule
        \end{tabular}
    }
\end{table}

\vspace{-3mm}
\subsection{Ablation and sensitivity studies}
\label{sec:ablation}
Table \ref{tab:ablation_and_sensitivity} shows two ablations (i.e., LOB conditioning and augmentation) and two sensitivity analyses (i.e., backbone choice and conditioning method) that highlight the effectiveness of TRADES design choice. 

\noindent\textbf{Ablation analyses.} We verify two of the hypotheses made in the method design: (1) how much the LOB conditioning part is necessary for the task and (2) how augmenting the feature vectors influences the performance. When we include LOB in the conditioning w.r.t. last orders only, TRADES has an average gain of $2.473$. When we augment the features through the MLPs in Fig. \ref{fig:architecture}, we gain an average of $1.980$ MAE absolute points. 
\begin{table}[!t]
\centering
\caption{Predictive score for the ablation (A) and sensitivity (S) analyses for two days of Tesla simulations.}
\label{tab:ablation_and_sensitivity}
\resizebox{.75\linewidth}{!}{%
    \begin{tabular}{clcc}
    \toprule
    & & \multicolumn{2}{c}{ \textbf{Predictive Score}$\mathbf{\downarrow}$} \\
    \cmidrule(lr){3-4} 
    & {\textbf{Method}} & {\textbf{ 29/01}} & { \textbf{30/01}}\\
    \midrule
    \multirow{2}{*}{A} & TRADES w/o LOB & 2.642 & 4.728  \\
    & TRADES w/o Aug. & 1.442 & 4.942\\
    \midrule
    \multirow{2}{*}{S} & LSTM backbone & 8.391 & 6.153  \\
    &TRADES w/ CA & 11.90 & 4.891  \\
    \midrule
    & TRADES (orig.) & \textbf{1.336} & \textbf{1.089} \\
    \bottomrule
    \end{tabular}
    }
\end{table}

\noindent\textbf{Sensitivity analyses.} We aim to verify whether the complexity of the backbone -- i.e., the transformer in TRADES -- is needed after all. Therefore, we replace the transformer backbone with an LSTM, leaving the augmentation and conditioning invariant. The table shows that the performances degrade by an average of $6.06$ absolute points. This is expected since transformers directly access all other steps in the sequence via self-attention, which theoretically leaves no room for information loss that occurs in LSTMs. 
Recall that we concatenate the past orders and the LOB snapshots, after the augmentation, into a single tensor and use it to condition the diffusion model -- i.e., TRADES (Orig.). 
We also tried a cross-attention (CA) conditioning strategy, -- see TRADES w/ CA --  between the past orders and the LOB snapshots. TRADES w/ CA reports an average performance loss of $7.814$ absolute points w.r.t. the original architecture. Note that cross-attention limits the model's capability because the orders cannot attend to each other but only to LOB and vice-versa.

\section{Conclusion}\label{sec:conclusion}

We proposed a Transformer-based Denoising Diffusion Probabilistic Engine for LOB Simulations (TRADES) to generate realistic order flows conditioned on the current market state. We adapted the predictive score to verify the usefulness of the generated market data by training a prediction model on it and testing it on real data. We showed that TRADES can cover the real data distribution by 67\% on average and outperforms SoTA by $\times 3.27$ and $\times 3.48$ on Tesla and Intel. Furthermore, we demonstrated that TRADES correctly abides by many stylized facts used to evaluate the goodness of financial market simulation approaches. 
We release DeepMarket, a Python framework for market simulation with deep learning and TRADES-LOB, a synthetic LOB dataset composed of TRADES's generated market simulations.  We argue that TRADES-LOB and DeepMarket will have a positive impact on the research community, as almost no LOB data is freely available. 
We believe that TRADES is a viable market simulation strategy in controlled environments, and further tests must be performed to have a mature evaluation trading strategy protocol. 
A promising experiment could involve training a larger model across various stocks, eliminating the need for an individual model for each stock.
\begin{ack}
This work was supported by a grant from Lazio Region, FESR Lazio 2021-2027, project @HOME (F89J23001050007 CUP B83C23006240002) and by the project “E-DAI”:  Piano Operativo Salute (POS) 2014-2020, CUP: B83C22004150001.
\end{ack}


\bibliography{bibliography}
\end{document}